\documentclass[journal]{IEEEtran}
\usepackage{cite}
\usepackage{graphicx}
\usepackage{float}
\usepackage{comment}
\usepackage{caption}
\usepackage{subcaption}
\usepackage{amsmath}
\usepackage{hyperref}
\usepackage{soul}
\usepackage[dvipsnames]{xcolor}
\ifCLASSINFOpdf
\else
\fi
\begin{document}
\title{Spatial Distribution of Data Capacity for the Reduction of Number of Repeaters in Ultra Long-Haul Links}
\author{Smaranika~Swain,~\IEEEmembership{Member,~OPTICA,}
Christian~Koefoed~Schou,~\IEEEmembership{Member,~OPTICA,}
 Metodi~Plamenov~Yankov,~\IEEEmembership{Member,~IEEE,}
Michael~Galili,~\IEEEmembership{Member,~IEEE,}
and~Leif~Katsuo~Oxenl{\o}we,~\IEEEmembership{Senior Member,~IEEE, Fellow,~OPTICA}
\thanks{All authors are with the Department
of Electrical and Photonics Engineering, Technical University of Denmark, Kongens, Lyngby, Denmark.}
\thanks{E-mail: lkox@dtu.dk, smasw@dtu.dk}
\thanks{Manuscript received xx}}
\markboth{Journal of \LaTeX\ Class Files,~Vol.~XX, No.~X, JANUARY~2024}
{Shell \MakeLowercase{\textit{et al.}}: Bare Demo of IEEEtran.cls for IEEE Journals}

\maketitle

\begin{abstract}
We present a novel method to reduce the number of repeaters and amplifiers in trans-oceanic
links by distributing a given data capacity in spatial channels. We analytically, numerically and experimentally demonstrate the principle and show that about 40\% of the repeaters can be omitted compared to a recently deployed cable. The method predicts that a single-fiber transmission link with 50 km amplifier spacing would be better off, repeater-wise, if the targeted single-fiber capacity is distributed in two fibers, each with an amplifier spacing of 150 km. In this scenario, one would thus only require 2/3 of the original number of amplifiers, and only 1/3 of the number of repeaters, housing the amplifiers. To test the principle of the proposed method, we  experimentally and numerically investigate a 6900-km long link with amplifier spacing of 50 and 150 km using a recirculating fiber transmission loop, and find that the result supports the analytical model and thus the proposed method. We then use this concept to analytically investigate a realistic $12$-fiber pair cable, and find that the same capacity could be distributed in $19$ fiber pairs requiring only about $56$\% of the original number of repeaters.
\end{abstract}

\IEEEpeerreviewmaketitle
\section{Introduction}
\IEEEPARstart{T}{he} demand for faster and ubiquitous Internet is consistently rising with more than 70\% of the global population being connected to the internet in 2023 \cite{Cisco1}. Transoceanic fiber cables form the backbone of the global Internet, and such cables have very specific demands and challenges, both in terms of design and in terms of deployment \cite{Grubb:18}. A recent report \cite{CableDamage} quantified the cable fault causes based on a large survey of installed submarine cables, and found that 32\% of cable faults stem from laying out the repeaters. Repeaters thus constitute a large component of the energy and cost budget of manufacturing, deployment, maintenance, and operation of subsea links (about twice the cost of transceivers \cite{Downie_Power_Conversion,Dar:2018}), and needing to keep personnel and deployment ships at sea to repair them during installation, makes them a liability and an investment risk that would be desirable to reduce. The authors have therefore developed a method for reducing the number of repeaters in a submarine cable considerably for a given data throughput. This could have impact on designs of future submarine cables, and components for such cables. 

Historically, submarine cable technology has undergone tremendous advancements \cite{Grubb:18, suboptic2019_ULL,suboptic2019_ofs} in the four main areas of capacity, complexity, cost, and power \cite{Sinkin, Downie}. From a power per bit perspective, it was found analytically for a single fiber link that amplifier spacings corresponding to 13 dB or less of amplifier gain is favorable \cite{Doran_analytic}. This implies installing a relatively high number of repeaters, and hence increasing the capital cost and risk of failures. Recently, space division multiplexing (SDM) has emerged as a proposition for modular capacity scaling
with reduced cost and energy per bit \cite{Sinkin_SDM,telecom, Bolshtyansky:20, suboptic2019_HFcountcable, MCF, MCF_2}, and the first generation of SDM-based optical submarine systems have been deployed \cite{DunantFieldTrial,16FP}, utilizing an increased number of fiber-pairs (FPs) with shared pump lasers for optical amplifiers
in repeaters. \cite{Downie_2022} reports on a detailed study of realistic SDM cables, including pump farming, in terms of achievable cable capacity per available power to support the amplifiers. It investigates a range of fiber losses and span lengths up to 120 km and suggests an optimum span loss around 10 dB. In \cite{Dar:2018}, a cost/bit perspective is introduced, and for submarine SDM cables, a
repeater spacing of 90 km is found analytically as an optimum. However, there has not been any numerical or experimental studies directly investigating the effect of greatly extended repeater spacing so far.

In this paper, we describe and demonstrate a novel method to reduce the number of repeaters in a link for a given data throughput, with amplifier spans extended beyond previous studies, with numerical and experimental validation. Unlike most approaches, where the optical power is maximised and then distributed into the available spatial channels, we suggest choosing a target capacity, and then distributing the data in multiple spatial channels. This \textit{data distribution strategy} is linked to the signal-to-noise ratio ($SNR$) through the Shannon-Hartley theorem, so that for a given target total capacity, the required $SNR$ in each spatially distributed channel, becomes significantly smaller than that for a single fiber channel. 
\begin{figure*}[t]
\begin{subfigure}{0.33\textwidth}
\includegraphics[width=7cm,keepaspectratio]{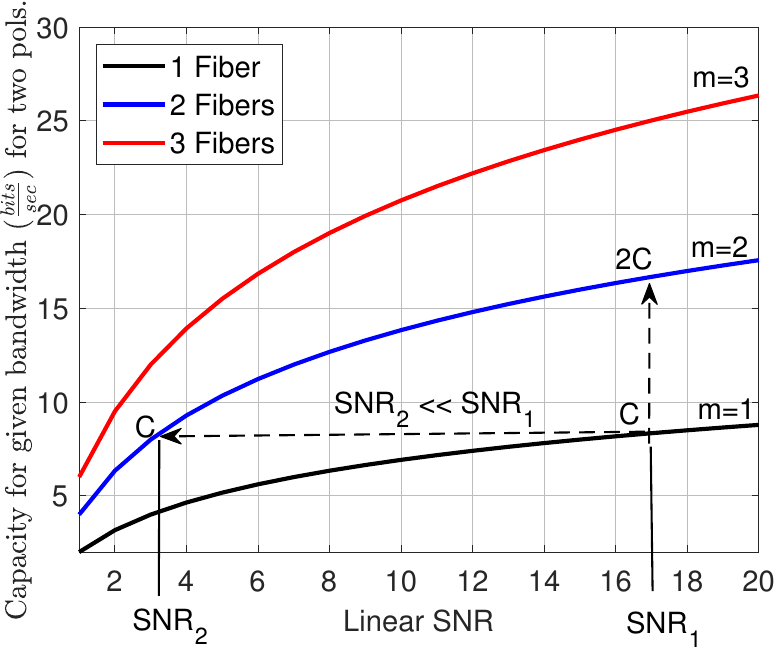} 
\caption{}
\label{fig:DF_idea1}
\end{subfigure}\hfill
\begin{subfigure}{0.56\textwidth}
\includegraphics[width=10cm, keepaspectratio]{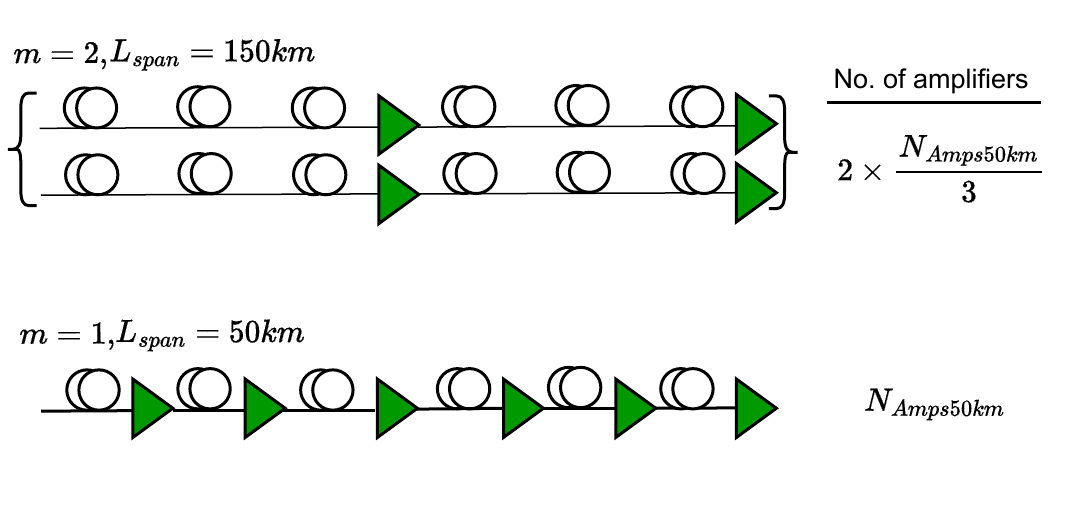}
\caption{}
\label{fig:DF_idea2}
\end{subfigure}\hfill
\caption{(a) Variation of capacity for a given bandwidth with SNR for two polarizations using Shannon’s channel capacity where $m$ is number of SDM fibers. (b) principle of the spatially distributed communication channel for $m=2$. }
\label{fig:DF_idea}
\end{figure*}
This low $SNR$ per fiber requirement may then in turn be utilized to distribute the repeaters at much greater distance than is customary in fiber links, significantly reducing the number of repeaters needed in a cable. 
We discussed the experimental implementation of this approach for a single optical carrier in \cite{CLEO_DF}. In this paper, we study the wavelength division multiplexed (WDM) case analytically, numerically and experimentally, and characterize the transmission  of five emulated WDM carriers with a central modulated channel in links with varying repeater spacing. Our analytical model suggests that increasing the repeater spacing from $50$ km to $150$ km by distributing the data capacity in two fibers instead of one would reduce the total number of Erbium-doped fiber amplifiers (EDFAs) in the link by 33\% for a $6900$-km link while maintaining the same data throughput. 
We thus investigate this scenario numerically and experimentally, and confirm the analytical performance for $50$ and $150$-km repeater spacing. In section II, we discuss the principle of spatial capacity distribution followed by an analytical formulation of required $SNR$ in the spatially distributed channels in Section III for a WDM system covering the C band. We discuss the numerical and experimental results demonstrating transmission with extended repeater spacing in Section IV and V respectively for five emulated WDM channels. In Section VI we consider the recently deployed Dunant cable \cite{DunantFieldTrial}, and analytically show that the same capacity could be achieved with only about $56$\% of the repeaters actually deployed. 

\section{Spatially distributed capacity}
\label{sec:II}
The basic principle of spatial capacity distribution is described in Fig. \ref{fig:DF_idea1}. Let us start by considering a single-fiber link designed to support a certain capacity ($C$) at the achievable signal-to-noise ratio ($SNR$). The relation between 
capacity and $SNR$ in a communication channel is given by the Shannon-Hartley theorem as $C = mB\log2(1 + SNR)$, where $m$ is the number of channels, here optical fibers, used to transmit the data, and $B$ is the available bandwidth and is set to unity for simplicity. As seen in Fig. \ref{fig:DF_idea1}, a certain $SNR_1$ per fiber corresponds to a given capacity, $C$. Adding more fibers multiplies the capacity ($2C, 3C$ etc.), corresponding to the spatial multiplexing case, i.e. moving vertically in Fig. \ref{fig:DF_idea1} to $m=2, 3$ etc at the original $SNR_1$ in each fiber. In our approach, we consider maintaining the capacity of the system, but distributing it in more fibers, thus going horizontally left in Fig. \ref{fig:DF_idea1} to lower $SNR$ in each fiber. Hence, distributing the same capacity in $2$ fibers, results in a lower required  $SNR$ ($SNR_2$) in each fiber. As seen in Fig \ref{fig:DF_idea1}, $SNR_2 << SNR_1$. This greatly reduced SNR requirement can be traded for different changes in the link design, e.g. cheaper amplifiers \cite{Dar:2018}, or, as suggested in this paper, extended repeater spacing ($L_{span}$), even far beyond the commonly used spacing in ultra-long-haul links. This principle is shown in Fig. \ref{fig:DF_idea2} for the case of $m = 2$, for realistic choice of parameters for amplifiers and fibers, corresponding to the analytical, numerical and experimental investigations, detailed in the following sections. Fig. \ref{fig:DF_idea2} schematically shows that the capacity of a single fiber can be distributed in two fibers, if the original single-fiber amplifier spacing of $50$ km is increased to $150$ km in the two fibers. Thus, one will only need $2/3$ of the total number of amplifiers from the original single-fiber case. The new amplifiers will of course need to provide higher gain to compensate for the loss in the $150$-km spans. Such gains of about $25$ dB are readily available in commercial amplifiers today e.g. \cite{Keopsys, Amonics}.  

In deployed submarine cables there are many fiber pairs, $N_{fp}$, so instead of considering a single-fiber converted into a dual-fiber data-distributed channel with 33\% reduction in amplifier numbers, one should consider how the number of repeaters (housing the amplifiers) in a submarine cable can be reduced by distributing the total capacity in more parallel fiber channels. Even for this more complex and realistic scenario, data distribution suggests large savings in repeaters and amplifiers, as discussed in Section \ref{sec:VI}.
\section{Analytical formulation}
\label{sec:III}
Fig. \ref{fig:DF_idea1} shows the capacity evolution for up to $3$ fibers. For the general case of $m$ fibers, we can derive the $SNR$ required in each of the $m$ channels ($SNR_m$) to achieve the same total capacity. Let us denote the capacity in the original single fiber as $C_1$ which is given by the Shannon-Hartley theorem as
\begin{equation}
    C_1=B\log_2 (1+SNR_1)
        \label{eq:III-1}
\end{equation} 
Distributing that same capacity in $m$ fibers, with $SNR_m$ in each fiber is written as 
\begin{equation}
    C_1=m B\log_2 (1+SNR_m)
    \label{eq:III-2}
\end{equation}
Equating (\ref{eq:III-1}) and (\ref{eq:III-2}) while assuming the same bandwidth, $B$, yields
\begin{align}
   \log_2(1+SNR_1)&=m \log_2(1+SNR_m) \\ \nonumber
  &=\log_2(1+SNR_m)^m
    \label{eq:III-3}
\end{align}
Cancelling $\log_2$ on both sides, and isolating $SNR_m$, reveals that 
\begin{equation}
SNR_m=\Big[1+SNR_1\Big]^{(1/m)}-1
\end{equation}
For large values of m, this approximates to
\begin{equation}
SNR_m \approx (1/m) \ln[1+SNR_1]
    \label{eq:III-4}
\end{equation}
The $SNR$ required in each of the $m$ fibers is thus completely determined by the initial SNR, i.e., $SNR_1$, saturates for large $m$.
\begin{figure}[t]
    \centering
\includegraphics[width=\linewidth,keepaspectratio]{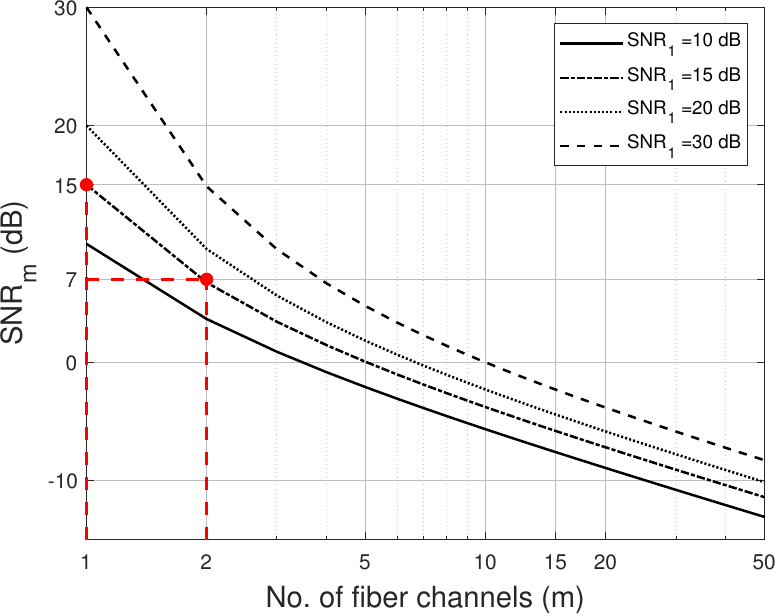}
    \caption{Dependence of $SNR_m$ on spatial channels, when distributing a given capacity in $m$ fibers, for different initial ($m=1$) $SNR$ values. The higher the initial $SNR$, the larger the drop in $SNR$ when distributing it in more fibers}
    \label{fig:SNRvsm}
\end{figure}
Note that, $SNR_m$ and $SNR_1$ represent the total available SNR of the system including all sources of noise and will be derived from the specific communication channel considered.
Fig. \ref{fig:SNRvsm} shows the evolution of $SNR_m$ (per fiber $SNR$) with respect to number of fibers, $m$, in which the initial total capacity is distributed, for different values of $SNR_1$. It shows that if the initial $SNR$ at $m=1$ (crossing with the y-axis) is high, a very steep drop in $SNR_m$ is observed for the first $2-5$ spatial channels the data is distributed in, and for $m>10$, the $SNR_m$ approaches a $1/m$ dependence, linear on the dB-scale in Fig.2. When the initial $SNR$ is quite low, the drop in $SNR$ requirement for $m>1$ is modest. 
\begin{figure}[t]
    \centering
\includegraphics[width=\linewidth,keepaspectratio]{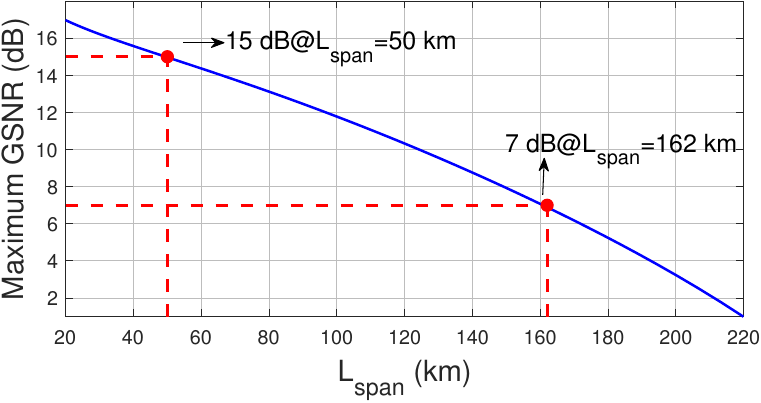}
    \caption{Variation of maximum $GSNR$ for varying amplifier spacing $L_{span}$}
    \label{fig:SNRvsLsapn}
\end{figure}
This indicates that a large saving in required $SNR$ can be capitalized on by distributing capacity in a few more spatial channels, if the initial $SNR$ is large. 
For instance, much lower $SNR$ may come about by increasing the spacing between the optical amplifiers in the link, as
suggested in Fig. \ref{fig:DF_idea2}. To determine the amplifier spacing from $SNR_m$, we set the amplifier gain ($G$) to exactly compensate for the transmission loss in the fiber span i.e.
\begin{align}
    G = \frac{1-\frac{P_{ASE}}{P_{in}}}{A}
\end{align}
where $P_{ASE}$ is the power of the amplified spontaneous emission induced by the amplifiers in the center channel. $P_{in}$ is the launch power in each span and can be seen as the signal power at the first span, and $A$ is the loss of each span and can be calculated as $A = \exp(-\alpha L_{span})$.
where $\alpha$ is the fiber loss coefficient in 1/km and $L_{span}$ is the distance between amplifiers in km. This looks different than a usual gain due to the introduction of the droop effect \cite{Downie_Droop} which takes into account the amplification of ASE power in the signal bandwidth.

The generalized SNR ($GSNR$), which includes impairments due to the fiber nonlinear interactions, amplified spontaneous emission and droop effect, can be calculated using the Gaussian noise (GN) model \cite{GNmodel,Downie,Downie_Droop} for different amplifier spacings as
\begin{align}
    GSNR = \frac{P_{s}}{P_{ASE}+P_{NL}} = \frac{\eta^{N_{sp}}}{1-\eta^{N_{sp}}}
 \label{eq:GSNR}  
\end{align}
where ${P_s}$ represents the signal power of the central channel, $P_{NL}$ is the power of the nonlinear distortion, $N_{sp}$ is the number of spans, and where $\eta$ can be calculated as follows:
\begin{align}
    \eta = 1-\frac{P_{ASE}+ P_{NL}}{P_{in}}
\end{align}
This model only works for the constant output power mode of the amplifiers and therefore $P_{in}$ should be the same for all spans. Note that penalties corresponding to non-ideal transceiver characteristics, non-ideal filtering, and quantization errors are not included in this model. 
Power corresponding to the noise due to amplified spontaneous emission is represented as
\begin{align}
\label{eq:P_ASE}
    P_{ASE} = 2hf_0n_{sp}(G-1)R_s
\end{align}
where $h$ is Planck's constant, $R_s$ is the symbol rate, $n_{sp}$ is the spontaneous emission factor, and $f_0$ is the frequency of the central carrier. The span length is thus included here through the gain and the number of spans. 
\begin{figure}[t]
    \centering
\includegraphics[width=\linewidth,keepaspectratio]{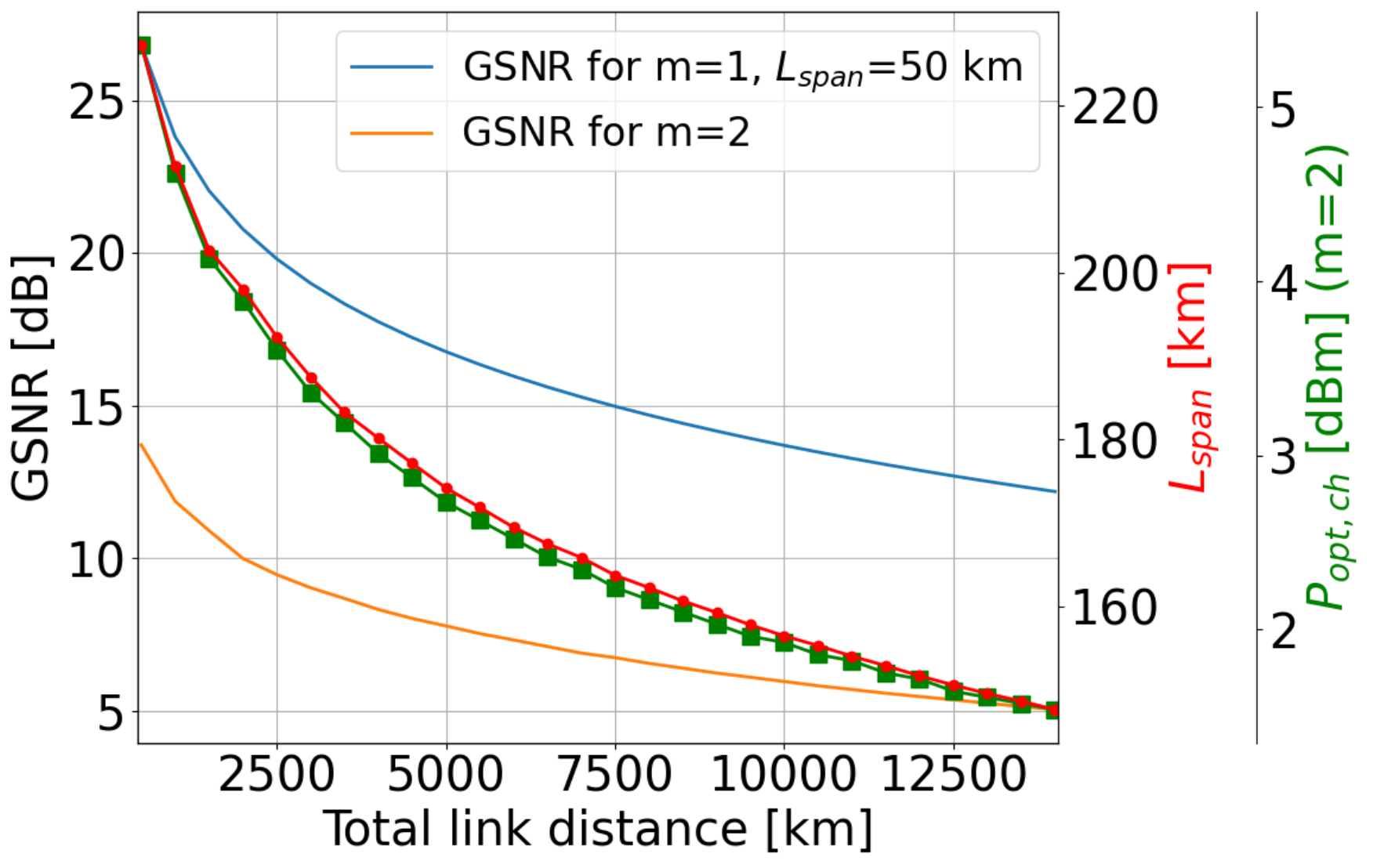}
    \caption{Variation of GSNR with total link distance for $m=1, L_{span}=50$ km and $m=2$ in the left y-axis. Variation of optimum span lengths for $m=2$ and corresponding optimum launch power per channel with total link distance in the right y-axis. }
    \label{fig:LspanvsL}
\end{figure}
The nonlinear noise contribution, $P_{NL}$, can be represented as \cite{GNmodel} 
\begin{align}
    & P_{NL} = \nonumber \\ 
    &\frac{8}{27}\gamma^2 L_{eff}^2 \frac{\text{asinh}\left( \frac{\pi^2}{2}|\beta_2|L_{eff_a}B_{WDM}^2 \right)}{\pi|\beta_2|L_{eff_a}}{\Big(\frac{P_{ch}}{B_{WDM}}\Big)^3}R_s
    \label{eq:beta_nonlinear}   
\end{align}
where $\gamma$ is the fiber nonlinearity coefficient and $\beta_2$ is the dispersion coefficient. The span effective length is denoted as $L_{eff}$ and is defined as $L_{eff} = \frac{1-\exp(-\alpha L_{span})}{\alpha}$ where $\alpha$ is the linear fiber loss coefficient. The asymptotic effective length is defined as $L_{eff_a} = \frac{1}{\alpha}$. $B_{WDM}$ is the combined bandwidth of all the WDM channels in each fiber. Thus the amplifier spacing is present in both the linear and the nonlinear part of the noise contributions. 
\begin{figure*}[t]
\centering
\includegraphics[width=\linewidth, keepaspectratio]{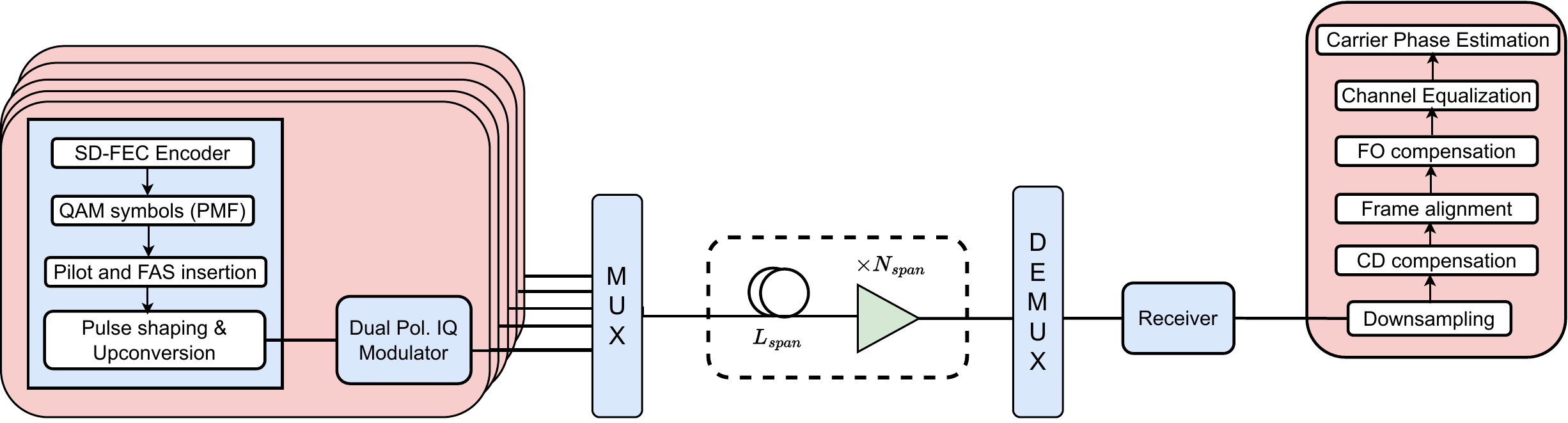}
\caption{Schematic of the simulation setup for 50 or 150 km repeater spacing for 5 WDM channels.}
\label{fig:sim_setup}
\end{figure*}
Furthermore, the optimal signal launch power into each span, $P_s$ grows for larger spacing, when operating at maximum $GSNR$. This will have a direct influence on the number of amplifiers a submarine cable can sustain within its limited power availability, as we will see in Section VI. 

The $GSNR$ for a single fiber as a function of amplifier spacing can now be calculated, corresponding to a specific capacity for given fiber nonlinearity, loss, amplifier noise figure, launch power, bandwidth etc. 
We consider a test system of length $6900$ km with inline EDFAs of noise figure $4.5$ dB supporting transmission of $117$ WDM channels modulated at $32$ GBaud covering a total signal bandwidth of $3.7$ THz. The properties of the optical fiber channel considered corresponds to that of TeraWave® SCUBA Ocean Optical Fiber with fiber loss ($\alpha)=0.155 $ dB/km, nonlinear coefficient ($\gamma)=0.715$ $W^{^-1}km^{-1}$, and dispersion coefficient $(D)=22$ ps/nm/km.
For each amplifier spacing, the maximum achievable $GSNR$, limited by nonlinearities, is identified, and this is plotted in Fig. \ref{fig:SNRvsLsapn} as a function of span length. This corresponds to the case of always driving the link at highest achievable $GSNR$. From Fig. \ref{fig:SNRvsLsapn}, one may now read out the $GSNR$ for a given amplifier spacing in a single fiber, which will correspond to a certain capacity for the $m = 1$ case, and by referring to Fig. \ref{fig:SNRvsm}, one can then read out the required $GSNR$ for the $m > 2$ cases, maintaining the same capacity, and then revert back to Fig. \ref{fig:SNRvsLsapn} to determine which increased amplifier spacing is allowed by this lower required $GSNR_m$. If one distributes the same capacity in $3$ fibers, the $GSNR$ in each fiber becomes even lower, corresponding to an even larger repeater spacing. But now there are $3$ fibers, and the savings on total number of amplifiers is smaller. Note that, for a submarine cable with $N_{fp}$ number of fiber pairs, a repeater station houses $2 \times N_{fp}$ number of amplifiers where the factor of $2$ corresponds to the two fibers in a fiber pair for bidirectional operation. Let us consider a cable with $N_{amp,1}$ amplifiers spaced $L_1$ km apart. In order to obtain a reduction in number of amplifiers, by spatially distributing in $m$ fibers, we need $\frac{N_{amp,m}}{N_{amp,1}}=\frac{m}{L_m/L_1}<1$, where $L_m$ represents amplifier spacings for different values of $m$. In other words, $L_m > mL_1$, needs to be achieved, which will only be sustained up to a certain value of $m$, after which the number of amplifiers will increase again. However, the number of repeaters may still be reduced by distributing the given data to more fiber pairs with increased repeater spacings (see Section VI), which may be desirable to reduce the risk of cable faults.  

Fig. \ref{fig:SNRvsLsapn} shows that the $GSNR$ gradually decreases for longer span lengths. If one now takes the $m=1$ case of $50$-km spacing in Fig. \ref{fig:SNRvsLsapn}, a $GSNR$ of about $15$ dB is found. From Fig. \ref{fig:SNRvsm}, we then see that going from $m=1$ to $m=2$ fibers, the required $GSNR$ drops from about $15$ dB to about $7$ dB in each of the two fibers. This means that the spacing in each of the two fibers can now be increased to $160$ km, as seen from Fig. \ref{fig:SNRvsLsapn}. Therefore, even if there are two fibers, since the spacing is more than $3$ times the original, one will overall need fewer amplifiers. Two recently deployed transoceanic cables Marea \cite{Marea} and the AEC2 \cite{telecom}  involve span lengths of 56 km and 54 km respectively. 

For the case of two parallel fibers ($m=2$), we performed further analytical simulations to evaluate the dependence of span length on total link distance. The left y-axis of Fig. \ref{fig:LspanvsL} shows the variation of $GSNR$ with total link distance for (i) a single fiber ($m=1$) with span length ($L_{span}$) of $50$ km and for (ii) two fibers ($m=2$) for different analytically estimated span lengths. The corresponding span lengths and optimum launch power per channel are shown in the right y-axis of Fig. \ref{fig:LspanvsL}. The optimum span lengths for 500 km, 7000 km (test case in the manuscript) and 14000 km links are found to be 227 km, 162 km and 144 km respectively which shows that for spatial data distribution using the proposed method the span length increases with shorter total link lengths.
As discussed  further in  Section VI,  one does not have to start with a single fiber, but could consider a full cable with  e.g. $8$ fiber pairs, and  consider  these $8$ fiber pairs as the $m=1$ case,  and then distribute the available capacity from these $8$ fibers at a given spacing, in more fibers with increased spacing, starting with $m=9/8, 10/8, 11/8$ etc.  In other words, for several fiber pairs, $m$ does not have to be an integer. In Section VI, we start out with $m=1=12/12$, and then distribute the capacity in more fiber pairs. 
In the following sections, we characterize the data distribution concept numerically and experimentally to confirm the theoretical predictions.   
\section{Numerical simulations}
\label{sec:IV}
We perform numerical simulations to study the effect of increasing repeater spacing on the system throughput. The schematic of the simulation setup is shown in Fig. \ref{fig:sim_setup}. Five wavelength division multiplexed (WDM) signals are simulated at $50$ GHz channel spacing in two polarizations. A random symbol sequence at $32$ GBaud is  pulse shaped using a square-root raised cosine filter with a roll-off of $0.01$ and modulated onto each WDM channel. The WDM signal is propagated through a multi-span EDFA-amplified link using the split step Fourier method (SSFM).
\begin{figure*}[ht]
\begin{subfigure}{0.33\textwidth}
\includegraphics[width=\linewidth, keepaspectratio]{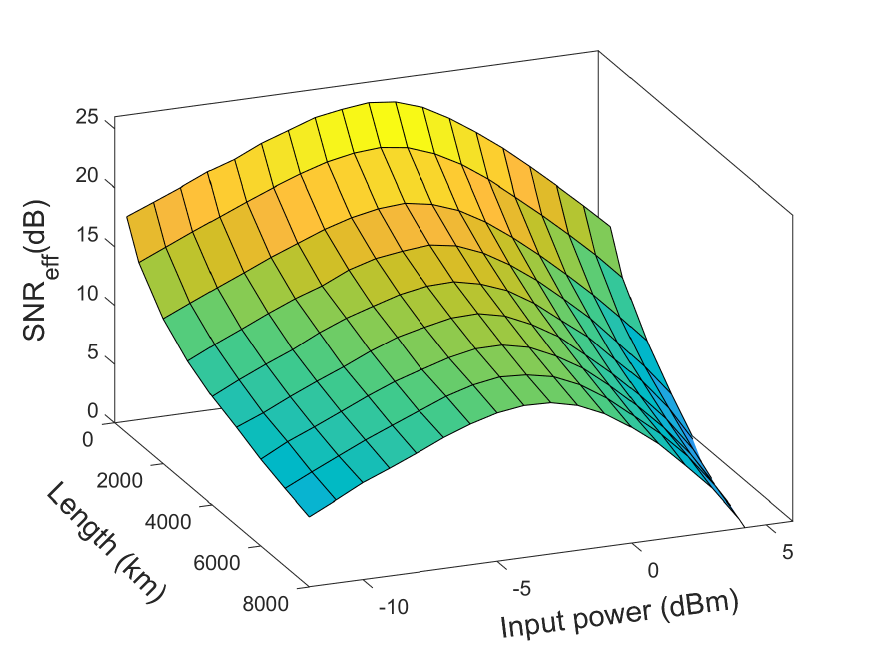}
\caption{}
\label{fig:sim_50}
\end{subfigure}\hfill
\begin{subfigure}{0.33\textwidth}
\includegraphics[width=\linewidth, keepaspectratio]{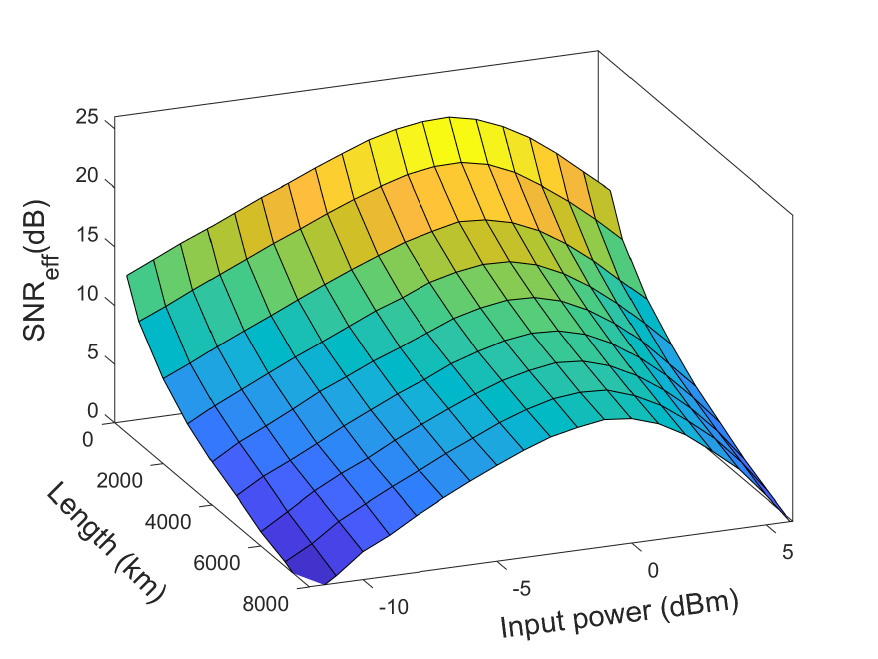}
\caption{}
\label{fig:sim_100}
\end{subfigure}
\begin{subfigure}{0.33\textwidth}
\includegraphics[width=\linewidth, keepaspectratio]{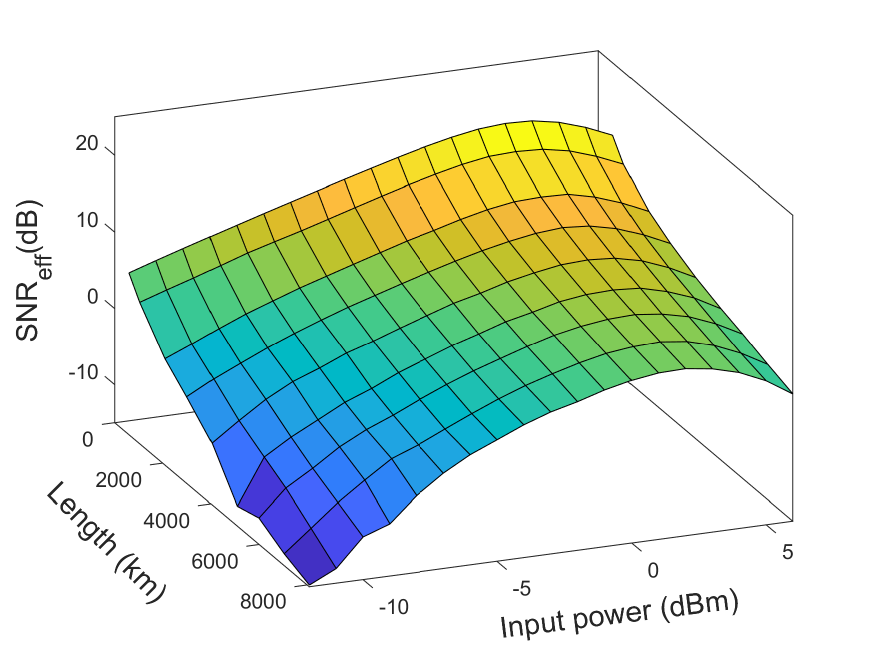}
\caption{}
\label{fig:sim_150}
\end{subfigure}
\begin{subfigure}{0.33\textwidth}
\includegraphics[width=\linewidth, keepaspectratio]{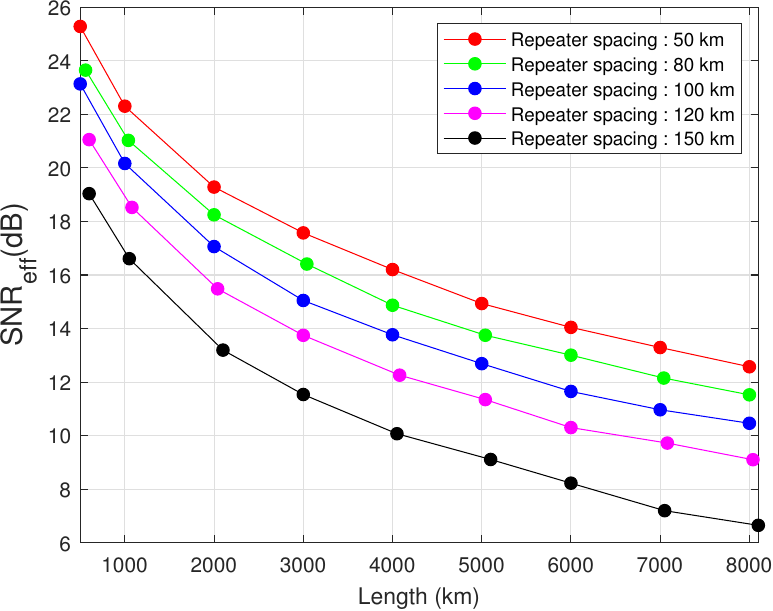}
\caption{}
\label{fig:SNREst_OFC}
\end{subfigure}\hfill
\begin{subfigure}{0.33\textwidth}
\includegraphics[width=\linewidth, keepaspectratio]{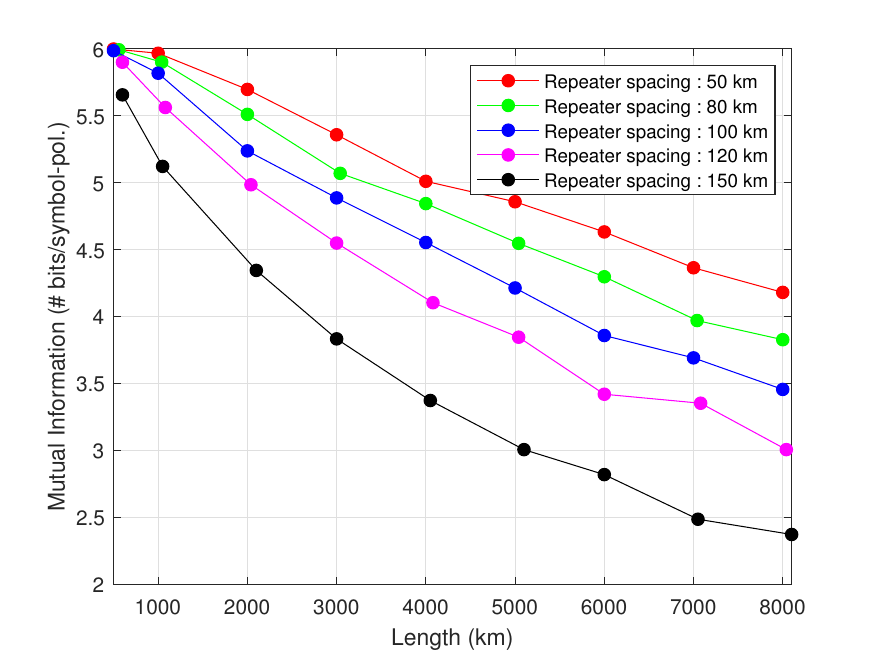}
\caption{}
\label{fig:MI_OFC}
\end{subfigure}\hfill
\begin{subfigure}{0.33\textwidth}
\includegraphics[width=\linewidth, keepaspectratio]{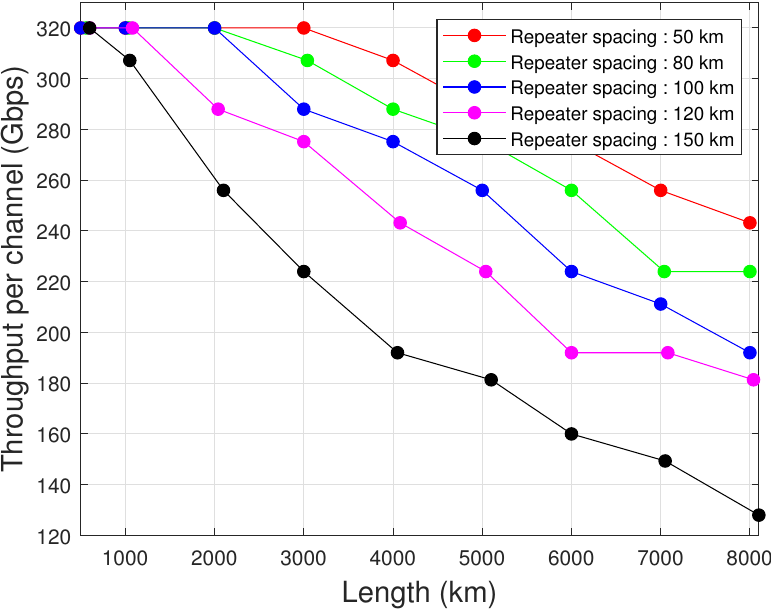}
\caption{}
\label{fig:throughput_OFC}
\end{subfigure}
\caption{Plots showing $SNR_{eff}$ of uniform PMF-64QAM signal simulated over SCUBA fiber as a function of fiber length for 5 WDM channels. (a) $SNR_{eff}$ at 50 km spacing as a function of total link length and for varying launch powers per channel. (b) $SNR_{eff}$ as in (a) but for 100 km spacing. (c) $SNR_{eff}$ for 150 km spacing. Plots showing (d) $SNR_{eff}$, (e) Mutual Information and (f) Throughput achieved at different link lengths for repeater spacings of 50 km, 80 km, 100 km, 120 km and 150 km at optimum launch power in each case.}
\end{figure*}
The properties of the optical fiber channel considered corresponds to that of TeraWave® SCUBA Ocean Optical Fiber, properties of which are mentioned in Section \ref{sec:III}. Span lengths ($L_{span}$) of $50$, $80$, $100$, $120$, and $150$ km are considered to understand the effect of gradual variation of repeater spacing on performance. For each span length, different total link lengths ($L_{total}$) varying from $500$ km to $8000$ km are considered to study practical transoceanic distances. 
\begin{figure*}[ht]
\begin{subfigure}{\textwidth}
\includegraphics[width=\linewidth, keepaspectratio]{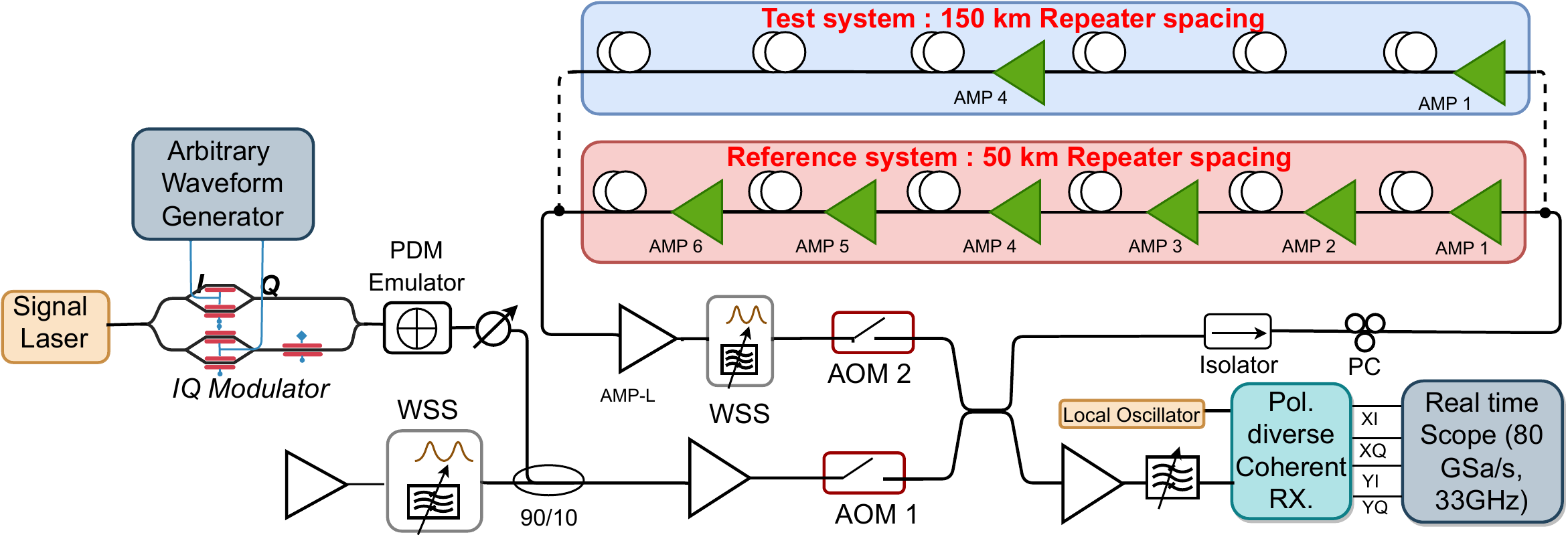}
\caption{}
\label{fig:Expt_setup}
\end{subfigure}\hfill
\begin{subfigure}{0.33\textwidth}
\includegraphics[width=\linewidth, keepaspectratio]{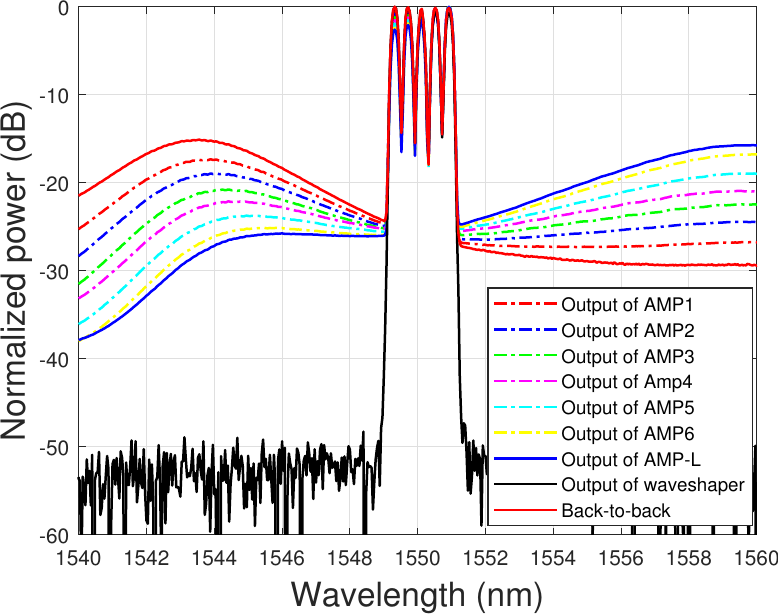}
\caption{}
\label{fig:Spectrum_Rep50}
\end{subfigure}\hfill
\begin{subfigure}{0.33\textwidth}
\includegraphics[width=\linewidth, keepaspectratio]{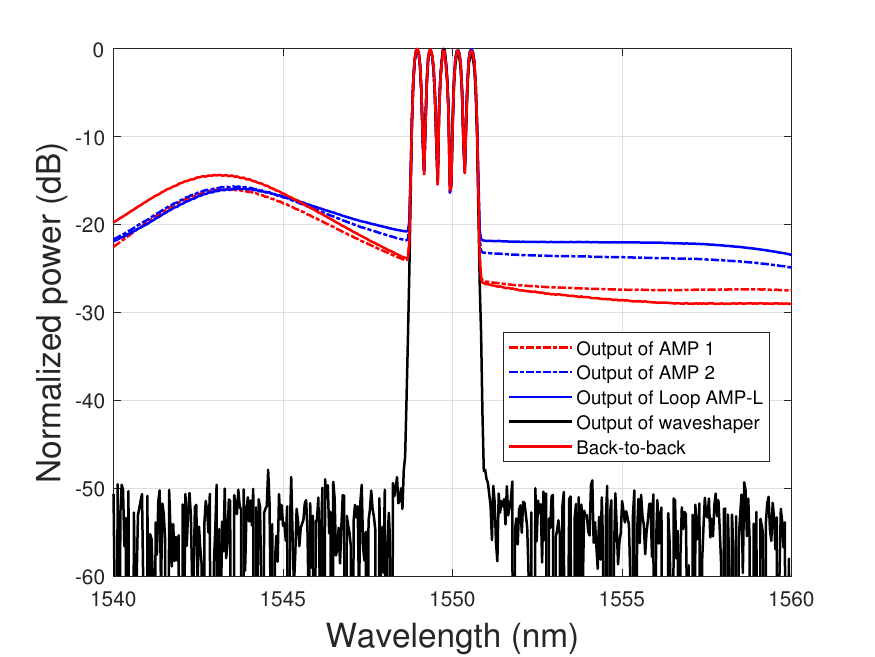}
\caption{}
\label{fig:Spectrum_Rep150}
\end{subfigure}
\begin{subfigure}{0.33\textwidth}
\includegraphics[width=\linewidth, keepaspectratio]{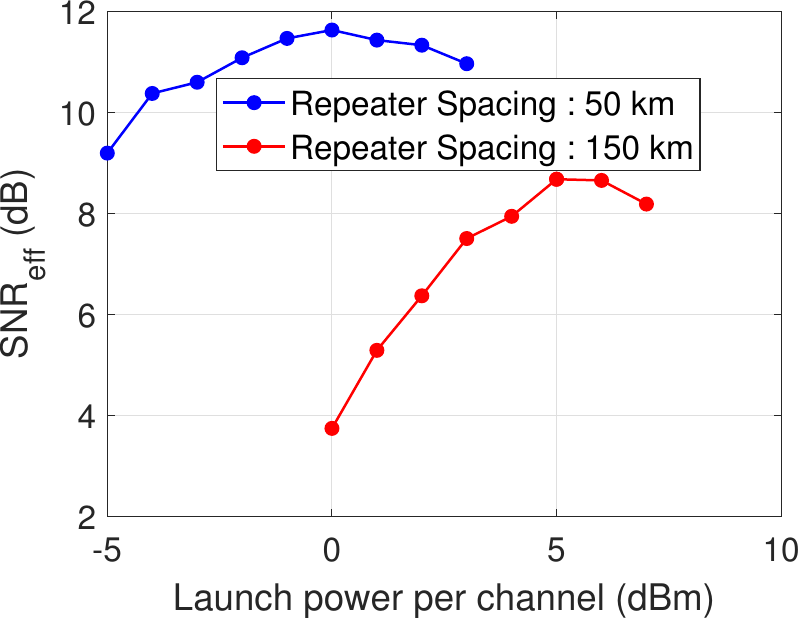}
\caption{}
\label{fig:LaunchPowerVar}
\end{subfigure}
\caption{(a) Schematic of the experimental setup with recirculating loop having 50 or 150 km repeater spacing for 5 WDM channels. Spectra of five WDM channels at different stages of the first loop turn for repeater spacing of (b) 50 km and (c) 150 km. (d) Dependence of effective SNR on launch power, average for all WDM channels, for different repeater spacings at a transmission distance of 3000 km.}
\end{figure*}

At the transmission side, the data sequence of length $K$ is
encoded by a forward error correction (FEC) code of length $N$. The FEC rate is then given by $R=\frac{K}{N}$. The input constellation symbols $x_k$
at time instant $k$ belong to the input alphabet $X$ of size $2^m$ and the received symbol sequence $y_k$ for one channel use belongs to the output alphabet $Y$. At the beginning of the transmitted signal sequence, a Zadoff-Chu sequence of length $101$ is inserted for frame alignment. This is followed by a training sequence ($x_{train}$). The training symbols have the same distribution as the payload and is known to the receiver for performance measurements. Pilot symbols of $4QAM$ modulation are inserted periodically in the sequence of transmitted symbols. Mutual information (MI) is then defined as  
\begin{equation}
    MI=\mathit{H}(\mathit{X})-\mathit{H}(\mathit{X|Y})
\end{equation}
where $H(X)$ represents the entropy of $X$ and $H(X|Y)$ represents the conditional entropy of $X$ when $Y$
has been observed. The entropy $H(X)$
represents an upper limit to the information rate and is
defined as $H(X) =-\sum_{x \in X}p_x \log_2p(x)$ where the probability of the input sequence $x$
is denoted $p(x)$. The conditional entropy instead describes how
much the uncertainty about $X$ is reduced by observing $Y$. The entropy,
and thus the information rate is ultimately bounded by the
size of the input alphabet i.e. $ H(X) \leq \log_2 |X|$.
Entropy matches with the size of input alphabet for the case of a uniform probability mass function (PMF).

In this work we use probabilistic constellation shaping (PCS) as a method of modifying the signal statistics to better match the
channel conditions by optimizing
their probability of occurrence in the
constellation. We implement PCS using the Maxwell-Boltzmann (MB) PMF obtained
using a constant composition distribution matcher (CCDM)
\cite{CCDM}. The MB PMF is a near-optimal family of distributions widely chosen for the fiber channel due to its simplicity \cite{MBPMF1,MBPMF2}. The rate of PCS is swept by varying the entropy $H(X)$ of
the MB PMF in the range $[2,6]$ bits/symbol with a step of $0.5$. 

 At the receiver side, conventional preliminary digital signal processing steps consisting of low-pass filtering, resampling, and frequency domain chromatic dispersion compensation are applied. The equalization involves a digital signal processing (DSP) chain consisting of a pilot-based, modulation-format-transparent, constant modulus algorithm using pilot symbols periodically interleaved with probabilistically constellation-shaped symbols.  The details of the equalization process is described in \cite{Metodi_PCS}. After equalization and frequency offset correction, the signal undergoes downsampling, phase noise tracking, bit-metric demapping and forward error correction (FEC) decoding. 
Transmission performance is evaluated in terms of effective signal-to-noise ratio ($SNR_{eff}$), mutual information ($MI$) and error-free data throughput. Effective SNR is defined as
\begin{equation}
SNR_{eff}=10\log_{10}\Big(\mathbf{E}\Big[\frac{|x_{train_{k}}|^2}{|\hat{y}_{train_{eq-k}}-x_{train_{k}}|^2}\Big]\Big)
\end{equation}
where $x_{train_{k}}$ is the training sequence, $\hat{y}_{train_{k}}$ corresponds to the equalized training sequence corresponding to the $k^{th}$ time instant and $\mathbf{E}[.]$ represents the expectation operation. 
Error-free performance is assumed
when the bit error rate (BER) after LDPC decoding is $< 10^{-5}$. Throughput is calculated at distances where the target data rate is transmitted without errors. Thus, throughput depends on the FEC rate and the entropy of the probabilistically shaped signal. It is defined as 
\begin{equation}
Throughput = R_s  \times \Big( H(X)- m \cdot (1- R) \Big) \times 2
\end{equation}
where $R_s$ represents baud rate, $m$ represents the number of bits/symbol and the factor of $2$ corresponds to polarization multiplexing.

Rate adaptivity is the key to achieving error-free performance in an $SNR$-limited system with varying $SNR$. In this work, we achieve rate adaptivity by sweeping (a) the modulation format ($16-$ and $64-$QAM), (b) entropy of the Maxwell-Boltzmann PMF and (c) coding overheads (OHs) (20\%, 25\% and 33\% low-density parity check (LDPC) codes from DVB-S standard \cite{LDPC}). For a certain combination of $L_{span}$ and $L_{total}$ and input PMF, the total power launched from the amplifier at the beginning of each span is swept and $SNR_{eff}$, $MI$, and information rate are evaluated at the optimal launch power. Fig. \ref{fig:sim_50}, \ref{fig:sim_100}, and \ref{fig:sim_150} show the corresponding surface plots of $SNR_{eff}$ with respect to input power and total link length for repeater spacings of 50 km, 100 km, and 150 km respectively with uniform PMF of a $64$-QAM constellation. It is inferred from Fig. \ref{fig:sim_50}, \ref{fig:sim_100}, and \ref{fig:sim_150} that for a given total link length, the optimal per channel launch power increases from $-2$ dBm in case of $50$ km repeater spacing to $3$ dBm in case of $150$ km repeater spacing. This is expected since an increase in span length implies an increase in nonlinear power threshold.

The optimal launch power is similarly obtained for different values of entropy ($H(X)$) of the Maxwell-Boltzmann PMF for each combination of total link length and repeater spacing. The effective SNR, mutual information and throughput achieved at different link lengths for the optimal combination of modulation format, input PMF and coding overhead that maximizes the mutual information at the receiver are evaluated and shown in Fig. \ref{fig:SNREst_OFC}, \ref{fig:MI_OFC}, and \ref{fig:throughput_OFC} respectively for repeater spacings of $50$, $80$, $100$, $120$, and $150$ km. It is observed that using a short repeater spacing of $50$ km, we achieve an MI of $4.18$ bits/symbol-pol at $8000$ km at an effective SNR of $12.5$ dB using $\frac{8000}{50}=160$ repeaters resulting in a throughput of $~250$ Gbps per WDM channel. In this case, the optimal combination is 64QAM, FEC-OH $= 25$\% and a PMF with entropy of $5$ bits/symbol/pol. However, if the repeater spacing is $150$ km, the throughput drops to $128$ Gbps at an effective SNR of $6.6$ dB using only $\frac{1}{3}$ ($\frac{8100 km}{150 km}=54$) of the repeaters. In this case, the optimal combination is 16QAM, FEC-OH $= 33$\%, and a PMF with entropy of $3$ bits/symbol/pol. The reduction in throughput by almost a factor of $2$ means that the addition of a second fiber pair would more than recover
the original throughput while at the same time yielding a $33$\% reduction in total number of repeaters. 
\section{Experimental Demonstration}
\label{sec:V}
To experimentally characterize the predictions of the data distribution concept, we use the experimental setup schematically shown in Fig. \ref{fig:Expt_setup} using $5$ emulated WDM channels and amplifier spacings of $50$ and $150$ km. The signal-under-test is modulated at 32 GBaud with a root-raised cosine pulse shape with $0.01$ roll-off. The in-phase ($I$) and quadrature-phase ($Q$) data generated by means of a $64$ GSa/s arbitrary waveform generator are modulated using an I/Q modulator onto an external cavity laser at $193.4$ THz with $\approx$ $10$ kHz linewidth. The modulated signal is then passed through a polarization-division multiplexing (PDM) emulator with a $1887$-symbols delay to ensure decorrelation between polarizations. The PDM signal is placed centrally among four wavelength division multiplexed (WDM) channels emulated by shaping amplified spontaneous emission (ASE) noise from an amplifier to emulate five WDM channels with 50 GHz spacing centered around $193.4$ THz. The WDM signal is amplified and applied to a recirculating loop. The recirculating loop is implemented using two acousto-optic modulators (AOMs)  used as optical switches. The transmission signal is passed through the first AOM (AOM 1) followed by a $2 \times 2$ optical coupler. One output of the coupler is connected to the transmission system using an isolator and the other output is connected to the receiver. A polarization
controller (PC) is used after the isolator to reduce the effect of any polarization
dependent gain accumulated in the loop. The transmission system consists of $300$ km of TeraWave® SCUBA Ocean Optical Fiber, kindly provided by OFS Denmark ApS. The transmission system is followed by an optical amplifier and a wavelength-selective switch (WSS) to amplify and shape the signal (flatten the spectrum) for the next recirculation. The signal is passed through a second AOM (AOM 2) and then to the other input port of the $2 \times 2$ coupler. AOM 1 and AOM 2 are triggered to allow for the desired number of round trips to emulate transoceanic link distances, after which the signal is passed to the receiver.

We consider two transmission systems in the recirculating loop, as shown in Fig. \ref{fig:Expt_setup}. The reference system involves an amplifier spacing of $50$ km, leading to the usage of $7$ identical amplifiers in the loop, out of which $6$ EDFAs (AMP \#1-AMP \#6) are used as amplifiers to compensate loss in the transmission fiber, while the last EDFA (AMP-L) in Fig. \ref{fig:Expt_setup} is used to compensate losses incurred in the recirculating loop. The test system involves a repeater spacing of $150$ km, leading to the usage of only $3$ amplifiers in the loop, two of which (AMP \#1 and AMP \#4) are used to compensate for fiber loss. The use of AMP-L is common to both link configurations. The spectra of the 5 WDM channels are shown in Fig. \ref{fig:Spectrum_Rep50} and Fig. \ref{fig:Spectrum_Rep150} at the output of the various EDFAs in the first turn of the loop, showing how the gain tilts towards longer wavelengths, revealing the need for gain flattening. 

 The recirculating signal experiences around $8$ dB of loss in each round-trip which corresponds to the insertion loss of AOM 2, the $2 \times 2$ coupler, the isolator and losses due to the process of spectral flattening using a wavelength-selective switch.  Only five WDM channels were chosen as a trade-off between target transmission distance and loss accumulated in equalization of gain tilt in each round trip. With low-loss equalization filters, the whole C-band could be supported and the output power required from the EDFAs is available from commercially available amplifiers. Note that, in a practical transmission link, the loss incurrred due to the loop does not exist. The gain of the EDFAs used for inline amplification is around $8.5$ dB and $25$ dB for repeater spacings of $50$ km and $150$ km respectively, which compensates the fiber transmission and various connector losses. Note that the EDFAs used in the experiment are standard inline amplifiers with $\approx 5$ dB noise figure.
At the receiver, the transmission signal is passed through a preamplifier after which the central WDM channel is filtered and detected using a polarization-diverse coherent receiver. 
The demodulated signal is digitized using an analog-to-digital converter and real-time oscilloscope operating at $80$ GSa/s. The oscilloscope is synchronized with the clock signal applied to the AOMs in order to capture digital data at desired propagated distances. The received digital samples are processed offline using rate-adaptive digital signal processing steps discussed in Section \ref{sec:IV} to extract SNR and post-FEC bit-error rate \cite{Metodi_PCS}. Fig. \ref{fig:LaunchPowerVar} shows the characterization of optimal launch power to achieve the maximum effective SNR ($SNR_{eff}$) for the 50 km and 150 km spacing cases for a total distance of 3000 km. As predicted by the numerical simulations, the launch power significantly increases at longer repeater spacing. About $5$ dB higher launch power is required for the $150$-km case with respect to the $50$-km case, agreeing with the simulations. The per channel launch power values shown in the x-axis of Fig. \ref{fig:LaunchPowerVar} are evaluated by dividing the total power at the output of each span amplifier by 5 (number of WDM channels). The total power is  measured using an optical power meter. Since there are no bandpass filters after each amplifier, the measured power values include out-of-band ASE, leading to the increase in optimal launch power compared to the numerical simulation results. The increased launch power for longer amplifier spacing will have an impact on the required electrical power in the cable, as discussed in Section \ref{sec:VI}. In order to compare the experimental results with simulations, we modify the simulation setup shown in Fig. \ref{fig:sim_setup} to emulate a recirculating loop for the cases of $L_{span}=50$ km and $150$ km for transmission of five WDM channels. 
After propagation in every $300$ km SCUBA fiber, we simulate a noise-loading ``loop'' EDFA to match the simulated $SNR_{eff}$ with that of the experimentally achieved $SNR_{eff}$ after one loop transmission. 
This allows us to include the effect of loss contributed by the recirculating loop.
\begin{figure}[t]
\centering
\includegraphics[width=\linewidth, keepaspectratio]{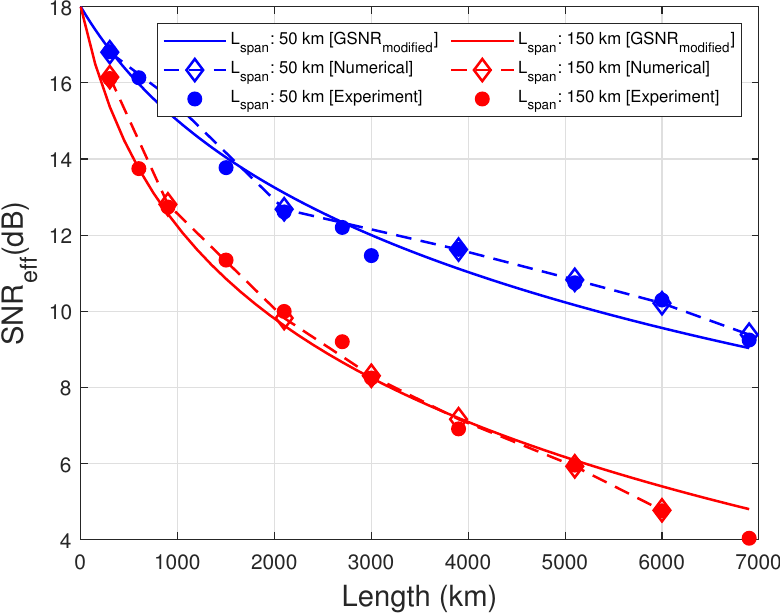}
\caption{Numerical and experimental variation of received $SNR_{eff}$ vs link length. Variation of $GSNR_{modified}$ using the analytical model given in (\ref{eq:GSNR_modified}) is shown as reference. The numerical simulation and the analytical model are modified to include loop losses.}
\label{fig:ExpCompare}
\end{figure}
In order to compare the experimental results with the analytical model using realistically achievable $SNR$ values from digital optical transmitters, we introduce a $GSNR$ limit in the back-to-back condition by modifying (\ref{eq:GSNR}) as follows
\begin{align}
     GSNR_{modified} = \frac{P_s}{P_{ASE}+P_{NL}+ \frac{P_s}{SNR_{B2B}} + P_N^{loop}}
     \label{eq:GSNR_modified}
\end{align}
where $P_N^{loop}$ represents the ASE noise contribution from the loop amplifier which is adjusted to match the loss incurred in one loop turn, $SNR_{B2B}$ is the maximum $GSNR$ that can be achieved in back-to-back condition. $SNR_{B2B}$ is now set to $18$ dB, which restricts the maximum SNR to the back-to-back value found in our setup. 
The numerically evaluated $SNR_{eff}$ and analytically evaluated $GSNR$ for five WDM channels including loop loss for link lengths varying from 300 km to 6900 km are shown to be in good agreement with the experimentally achieved $SNR_{eff}$ for repeater spacings of $50$ km and $150$ km as shown in Fig. \ref{fig:ExpCompare}.We observe that at 6900 km, $SNR_{eff}$ changes from about $10$ dB for a 50-km repeater spacing to about $5$ dB for a 150-km repeater spacing. This reduced SNR corresponds to the required SNR to support half of the total capacity as shown in Fig. \ref{fig:CapDualFiber}.

In Fig. \ref{fig:CapDualFiber}, the experimentally achieved throughput, i.e. the processed error-free received bit rate, is shown for repeater spacings of $50$ km and $150$ km for increasing link lengths. The modulation format, PMF of constellation, coding rate, and launch power are optimized for each link length to achieve the maximum throughput. As expected from the concept of spatial data distribution discussed in Section \ref{sec:II}, at 6900 km, the 150 km case yields a little higher than half the throughput of the 50 km case. 
When two fibers are employed with 150 km span length, one would thus achieve a little more than the original capacity from the 50-km span length case. It demonstrates clearly that, increasing repeater spacing reduces the per-fiber SNR, and hence per-fiber throughput, but maintains the total cable throughput by increasing the fiber count, and this happens while reducing the total number of repeaters. Note that 
\begin{figure}[t]
\centering
\includegraphics[width=\linewidth, keepaspectratio]{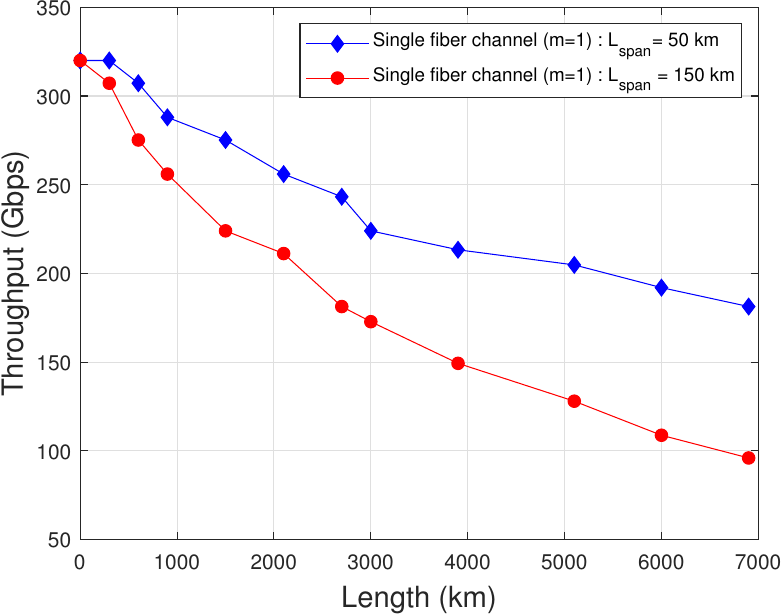}
\caption{Averaged experimental throughput per WDM channel
(Gbps) in a single fiber (red and blue solid lines) vs link lengths up to 6900 km for repeater
spacings of 50 km (diamond markers) and 150 km (lower trace with dot markers).}
\label{fig:CapDualFiber}
\end{figure}
in the analysis of this paper, we have simply assumed a direct proportionality of number of identical transceivers to number of fibers, i.e. $2 \times N_{fp}$ leads to twice the number of identical transceivers. 
However, speculating on the advancement in photonic integration, the transceiver technology for low-SNR transmission could potentially result in simpler transceivers.
\section{Analysis of electrical power constraints} 
\label{sec:VI}
Electrical power supplied from the ends of the submarine cable is a critical resource, especially in case of the emerging SDM technology where increased number of fiber pairs need to share the available electrical power. The idea of spatial distribution of capacity in two fibers (as discussed in Section \ref{sec:IV} and \ref{sec:V} of this work) can be generalized to $m$ fibers as analytically studied in Section \ref{sec:III}. However, this distribution strategy is strictly limited by the available electrical power in a submarine cable. Maximum single-ended cable voltage supplied in commissioned cables currently is $ 18$ kV \cite{18-20kV}, and often $ 15$ kV. The power available for each repeater amplifier can be described as \cite{Downie}:
\begin{equation}
\label{eq:P_avai}
    P_{available}=\frac{(V_{PFE})^2}{4N_{sp}LR_0N_{rep}}
\end{equation}
where $V_{PFE}$ is the voltage power feed, $N_{sp}$ is the number of spans in a single fiber channel, which is equivalent to the number of amplifiers. The total length of the cable is $L$ which has a resistive coefficient of $R_0$ $\Omega /km$. $N_{rep}$ represents the number of repeaters in the entire cable.
\begin{figure*}[ht]
\centering
\begin{subfigure}{.3\textwidth}
  \centering
  \includegraphics[width=\linewidth]{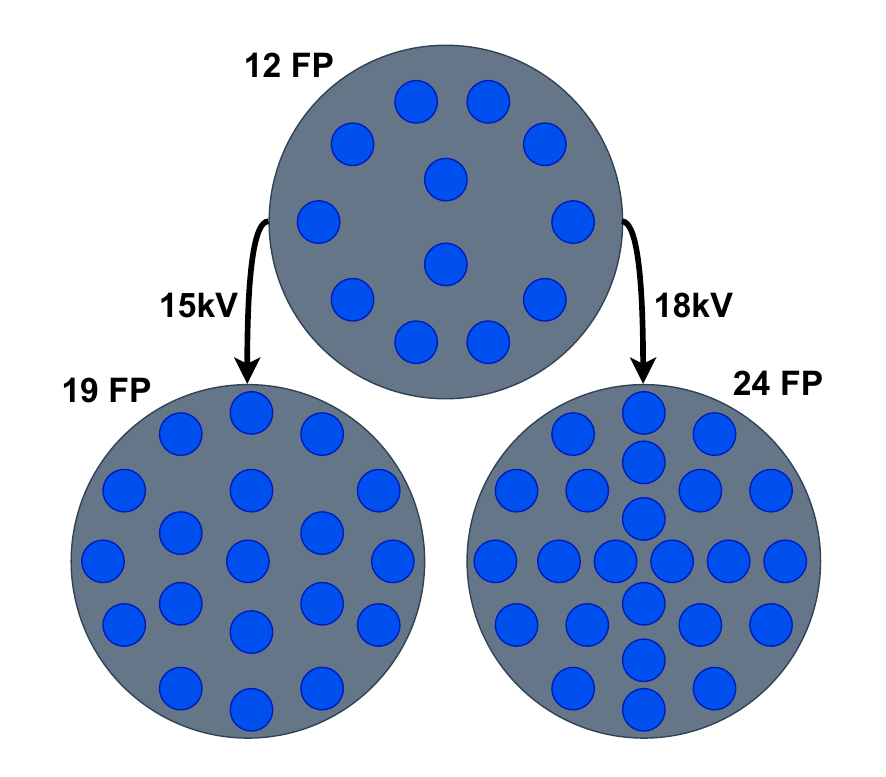}
\end{subfigure} \qquad
\begin{subfigure}{.65\textwidth}
  \centering
  \includegraphics[width=\linewidth, keepaspectratio]{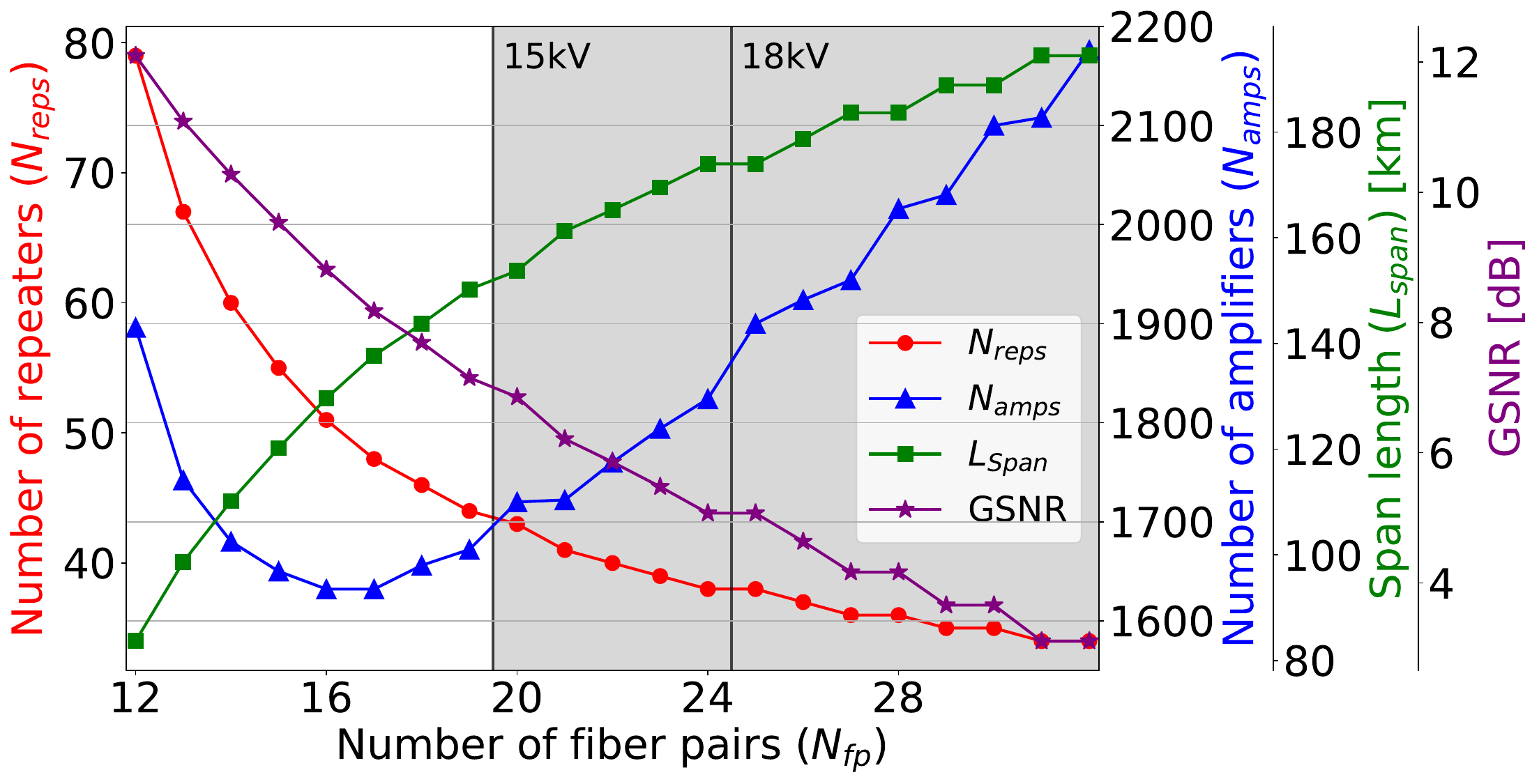}
\end{subfigure}%
\caption{\textbf{Left:} Comparing the Dunant cable with 12 fiber pairs to the maximum achievable fibers of 19 or 24 for $V_{PFE} =15$ and $18$ kV respectively using the model described in sec. \ref{sec:II} and \ref{sec:VI}.
\textbf{Right:}Required number of amplifiers and number of repeaters for a varying number of fiber pairs for a 6611km link, initially carrying 12 fiber pairs with a spacing of 84 km and capable of supporting a target throughput of 220 Tbit/s. The shaded region beyond 19 fiber pairs, corresponds to the cases where there is no longer enough power in the cable available to support running the fiber pairs at the optimal GSNR for $V_{PFE} = 15$ kV.}
\label{fig:DunantTest}
\end{figure*}
Since a repeater unit consists of $2N_{fp}$ number of amplifiers located every $L_{span}$ distance of the fiber cable, the power needed for each repeater can be found as
\begin{equation}
    \label{eq:P_req}
P_{repeater,required}=\frac{2N_{fp}N_{ch}P_{ch}}{(1-\epsilon)\eta}
\end{equation}
where $N_{fp}$ is the number of fiber pairs, $N_{ch}$ is the number of optical channels, $\epsilon$ is the fraction of the electrical power that is used for the control circuit, and $\eta$ is the electrical to optical power efficiency \cite{Downie, cable_power}. Note that, in (\ref{eq:P_req}) the electrical power is not proportional to the gain of the amplifier but the optical output power. Combining (\ref{eq:P_avai}) and (\ref{eq:P_req}) we find the maximum number of fibers that can be supported as:
\begin{align}
    N_{fp}&=\frac{P_{available}}{ P_{repeater,required}} \nonumber \\
    &=floor\Bigg[\frac{(1-\epsilon)\eta(V_{PFE})^2}{8N_{sp}LR_0N_{rep}N_{ch}P_{ch}}\Bigg]
\end{align}
Using these power assumptions we can consider a test case, the deployed $6611$-km Dunant SDM cable with $12$ fiber pairs and an average $84$ km repeater spacing \cite{DunantFieldTrial}. We use the analytical model discussed in Section \ref{sec:III} to evaluate the maximum number of fiber pairs allowed to achieve a target capacity, while utilizing the available electrical power of $15-18$ kV \cite{Downie}. 
Starting with 12 fiber pairs, at an average spacing of 84 km over a 6611 km link \cite{DunantFieldTrial}, yields a supported capacity of $220$ Tbit/s, assuming SCUBA fibers and transmission in the full C-band, using amplifiers with noise figures of $4.5$ dB and Nyquist pulse shaping. This capacity can now be distributed in more fibers, adding one at a time, and determining the repeater spacing, and optimizing the launch power to achieve the highest possible $GSNR$. Since subsea SDM cables mean the possibility to distribute the electrical powering towards more fibers in parallel, operating the system 1 or 2 dB below the nonlinear threshold, is more realistic. With the assumption that the amplifiers are operating at 2 dB below the nonlinear threshold, Fig. \ref{fig:DunantTest} shows how the cable characteristics evolve when distributing the data in more fiber pairs. The fluctuations of these curves can be explained by the number of repeaters, amplifiers, and fiber pairs only increasing in integers, which makes some designs more or less favorable when a specific capacity is required. 
We see that by distributing the same data in more and more fibers, the repeater numbers continue to decrease, as the repeater spacing continues to grow. We also see that the total number of amplifiers housed in the repeaters reach a minimum, after which adding more fibers will require more amplifiers in yet fewer repeaters. This could require considerable design changes of repeaters and be a practical challenge. More fundamentally, the power required to drive more amplifiers will set a hard limit. The shaded region in Fig. \ref{fig:DunantTest} marks where the power supplied to the cable is depleted, assuming the use of $15$ kV power supply. Therefore, under the constraints that each fiber is running at launch powers corresponding to 2 dB below the maximum $GSNR$, and the electrical power is finite, the data distribution concept suggests that the fewest sustainable repeaters are obtained by extending the repeater spacing from $84$ km to $150$ km, and simultaneously expanding the number of fiber pairs from $12$ to $19$. In doing so, the repeaters required would drop from $79$ to $44$, i.e. a drop of $44$\%, and the number of amplifiers in those repeaters would drop from $1896$ to $1672$ total number of amplifiers, i.e. a drop of $12$\%, all under realistic power supply conditions. As a $16$-fiber pair cable has been deployed \cite{16FP}, this design could be realistic, with optimised amplifier and cable designs in the future. For $V_{PFE}=18$ kV, the repeater count could be lowered even further, and up to $24$ FP could be supported while maintaining the capacity. The repeater count then drops to $38$ which is a $52$\% decrease, leading to a span length of $174$ km, while the number of total amplifiers is $1824$ which is a decrease of $4$\%. In this analysis, we operate at launch powers 2 dB below the maximum $GSNR$. If we were to operate at the maximum GSNR, the number of supported fiber pairs would only amount to 14.


\section{Conclusion}
We show that a single-fiber link with a 50-km repeater spacing may be replaced by a dual-fiber link with repeaters
spaced thrice the distance of the single-fiber case, saving about $33$\% of the total amplifiers. We analytically, numerically, and experimentally demonstrate this approach for reducing the number of repeaters in trans-oceanic links. This approach of spatial capacity distribution will require amplifiers with higher total output power, which are currently being investigated \cite{suboptic2019_repeater,suboptic2019_thermalmanagement}. We also find that for a realistic 12-fiber pair cable, the data distribution concept can offer a saving of about $44$\% of the required repeaters, by distributing the capacity in $19-24$ fiber pairs, and still stay within the available supply voltages for deployed cables. Saving such considerable numbers of repeaters could be beneficial in reducing breaking points in cables, and associated costs. With the advent of ultra-low loss fibers \cite{suboptic2019_ULL} , high fiber-count cables \cite{suboptic2019_HFcountcable}, deployment ready multicore fibers \cite{24FP}, and single mode fibers with reduced cladding diameters \cite{coating}, spatial parallelism is expected to play a major role in the evolution of submarine communication links. This work emphasises that the optimization of the combination of repeater spacing and number of fiber pairs could play a role in future subsea link design strategies.



\section*{Acknowledgment}
The authors would like to thank OFS Denmark Aps for the SCUBA optical fiber used in the experiments. This work is supported by the SPOC Centre (Silicon Photonics for Optical Communications) funded by the Danish National Research Foundation (DNRF123).
\bibliographystyle{unsrt}
\bibliography{Bib}
\begin{IEEEbiography}{Author1}
Biography
\end{IEEEbiography}

\begin{IEEEbiographynophoto}{Author2}
Biography
\end{IEEEbiographynophoto}

\begin{IEEEbiographynophoto}{Author3}
Biography
\end{IEEEbiographynophoto}

\end{document}